\documentclass{elsarticle}
\usepackage{graphicx}
\usepackage{subfigure}
\usepackage{lineno,hyperref}

\begin{document}

\title{$\gamma$-ray emission signals in the massive graviton mediated dark matter model}
\author[address1,address2]{Cun Zhang}

\author[address1,address2]{Ming-Yang Cui}

\author[address2]{Lei Feng  \corref{mycorrespondingauthor} }
\ead{fenglei@pmo.ac.cn}

\author[address2]{Yi-Zhong Fan \corref{mycorrespondingauthor}}
\ead{yzfan@pmo.ac.cn}

\author[address1]{Zhong-Zhou Ren
\corref{mycorrespondingauthor}}
\ead{zren@nju.edu.cn}

\cortext[mycorrespondingauthor]{Corresponding author}

\address[address1]{School of Physics, Nanjing University, Nanjing, 210092, China}
\address[address2]{Key Laboratory of Dark Matter and Space Astronomy, Purple Mountain Observatory, Chinese Academy of Sciences, Nanjing
210008, China}

\begin{abstract}
Dark matter may interact with Standard Model (SM) particle through the exchange of a
massive spin-2 graviton producing signals that can be detected. In this work we examine the $\gamma$-ray emission signals, including the line
emission and the continuous spectrum component in such a massive graviton-mediated dark matter model. The constraints of LHC data, dark matter relic density
 as well as the dark matter indirect detection data have been applied to narrow down the parameter space. We focus on the vector dark matter model which could produce detectable $\gamma$-ray line signal.
  It is found that the $\gamma$-ray line data is effective on constraining the model parameters and  the ongoing and upcoming space or ground-based $\gamma$-ray experiments can constrain the model further.
 As for the continuous $\gamma$-ray emission, the total effective annihilation cross section is $\sim 10^{-26}
{\rm cm^{3}~s^{-1}}$  except at the region where dark matter mass is around the graviton mass or half of it, which is consistent with current observational data and will be reliably probed by the upcoming CTA.
\end{abstract}
\maketitle

\section{Introduction}
Dark Matter (DM) consists of 26 percent of the total energy and 84 percent of the total mass of the current universe in $\Lambda$CDM cosmology \cite{plank}. So far the evidences of
DM are mainly from astronomical scale phenomena, such as rotation curve of galaxies \cite{galaxy}, cosmic microwave background (CMB)\cite{CMB1}, gravitational lensing\cite{lensing} and so on \cite{dark,dark1}. Several types of experiments are designed to detect the dark matter particles (i.e., the so-called direct, indirect and collider detection). The Large Hadron Collider (LHC) is
searching the dark matter particles events with different final states produced in hadron collisions \cite{LHCd1,LHCd2,LHCd3,LHCd4}. Indirect detections are searching the final particles of dark matter annihilation or decays in the cosmic ray data and the on-orbit experiments include for example Fermi-LAT \cite{Fermi,Fermi1,Fermi2}, AMS-02\cite{ams1,ams2,ams3}, Dark Matter Particle
Explorer (DAMPE \cite{dampe}) and  CALorimeteric Electron Telescope (CALET \cite{calet}). If dark matter could collide with nuclei, the recoiled nucleus can be detected directly. Such experiments are usually done in underground laboratory to avoid the interference of cosmic rays, such as LUX \cite{LUX}, Xenon \cite{xenon}, CDEX \cite{cdex} and PandaX-II \cite{panda}.

The dark matter particles are considered to be some unknown particles and their interaction can not be described by the Standard Model(SM). Various
dark matter particle candidates have been studied, such as, the weakly interacting massive particles (WIMPs) \cite{WIMP1,WIMP2,WIMP3}, axion \cite{axion1,axion2,axion3}, Kaluza-Klein (KK) particle \cite{kk1,kk2,kk3} and so on. In this work we study a simple model where the dark matter particles interact with  the transverse components of SM gauge bosons through a massive graviton mediator.
A massive graviton can present in different models, for example, the lightest one of KK gravitons
\cite{gravtity0,gravtity1,gravtity2} in the Randall-Sundrum (RS) model \cite{RS1,RS2} or Arkani-Hamed-Dimopoulos-Dvali (ADD) model \cite{add} and
the ghost-free massive graviton in  de Rham–Gabadadze–Tolley (dRGT) model \cite{spin2-1,spin2-2,spin2-3,spin2-4}. Such a kind of model was used in \cite{gravtity2} to interpret the 750 GeV diphoton excess suggested in 2015. Though the diphoton exceess is most-likely due to the statistical
fluctuation \cite{a750GeV,pp2aa}, the massive graviton mediated dark matter model is still attractive and should be further explored. In this work we try to constrain the parameter space of such a model and the mediator mass is set to be free.

 Lee {\it et al.} \cite{gravtity3} studied the astrophysical bounds on the KK graviton mediated dark matter where the effect of the whole KK tower is considered.
These authors constrained the 2-dimension parameter space (i.e, other parameters are fixed) using bounds for monochromatic photons, continuous $\gamma$-ray
and $\gamma$-ray boxes from Fermi-LAT and HESS.
The LHC has made a great efforts to search new
resonances in different channels ($W^+W^-$,$ZZ$,$e^+e^-$, and so on). Though no
significant resonance has been identified, the corresponding upper limits can be used to tightly constrain the model parameters. In the meantime, Fermi-LAT has collected more data and reported stronger bounds on dark matter induced $\gamma$-ray signal.
The main purpose of this work is to explore the parameter space allowed by the thermal relic density, current data $\gamma$-ray (including the line emission and the
continuous spectrum component) and LHC data (resonance in
different channels and monojet) in effective massive graviton mediated dark matter model. The vector dark matter model which could produce detectable $\gamma$-ray line signal, widely believed to be the smoking-gun signature of dark matter annihilation or decay, is focused. The detection
prospect of the dark matter-induced signals by the future experiments is also discussed.

This work is organized as follows: In Section II we briefly introduce the effective massive graviton mediated dark matter model, then we constrain the model parameters with the observational data and discuss the detection prospect of some ongoing and upcoming experiments. Our results are summarized in Section \ref{sec:summary}.

\section{Model and constraints}
\subsection{Model}\label{sec:Models}
In this work we study a simple model where the dark matter particles interact with SM gauge bosons through a massive graviton mediator. The effective interaction between the massive graviton $G^{\mu \nu}$ and other particle is given by
\begin{equation}
\mathcal{L}_{\rm int}=-\frac{c}{\Lambda}G^{\mu \nu}T_{\mu \nu},
\end{equation}
where $\frac{c}{\Lambda}$ is the full coupling constant, $\Lambda$ denotes the coupling scale which is set as 3 TeV in this work, $c$ is a dimensionless coupling constant and $T_{\mu \nu}$
is the energy-momentum tensor of the particle. It should be pointed out that we just discuss a phenomenal effective model here. Note that for spin $0 ~(1/2)$ dark matter, the dark matter annihilating into a pair of photons channel is d-wave (p-wave) suppressed, giving rise to a weak $\gamma$-ray line signal. While for spin-1 dark matter,
the channel is s-wave and the generated $\gamma$-ray line signal might be detectable. In this work we focus on the vector dark matter model.
The energy-momentum tensor of vector dark
matter is
\begin{eqnarray}
T^{V}_{\mu \nu}=&&\left[ \frac{1}{4} g_{\mu \nu} X^{\lambda \rho} X_{\lambda \rho} -X_{\mu \lambda}X^\lambda_\nu + m^2 _X \left( X_\mu
X^\nu -\frac{1}{2} g_{\mu \nu} X^\lambda X_\lambda \right) \right],
\end{eqnarray}
where $X_{\mu \nu}=\partial_\mu X_\nu -\partial_\nu X_\mu$ and $X_\mu$ is the field of the vector dark
matter, $m_X$ is the rest mass of the vector dark matter.

In the case of that
the graviton mainly interact with the transverse components of the gauge bosons which proposed in Ref.\cite{gravtity2}, the coupling with SM particle can be expressed as
\begin{eqnarray}
\mathcal{L}_{\rm{G-SM} }= && -\frac{1}{\Lambda}G^{\mu \nu} \left( \sum_{a=1}^{3} c_a(\frac{1}{4}g_{\mu \nu}F^{\lambda \rho}_a F_{\lambda \rho, a} -F_{\mu
\lambda,a}F^\lambda_{\nu,a}) \right)
\nonumber \\
= &&-\frac{1}{\Lambda}G^{\mu \nu} \left[ c_{\gamma \gamma} \left( \frac{1}{4} g_{\mu \nu} A^{\lambda \rho} A_{\lambda \rho} -A_{\mu
\lambda}A^\lambda_\nu \right)\right.\nonumber\\
 && + c_{Z \gamma} \left( \frac{1}{4} g_{\mu \nu} Z^{\lambda \rho} A_{\lambda \rho} -Z_{\mu \lambda}A^\lambda_\nu \right)
    +c_{Z Z} \left( \frac{1}{4} g_{\mu \nu} Z^{\lambda \rho} Z_{\lambda \rho} -Z_{\mu \lambda}Z^\lambda_\nu \right)\nonumber \\
&& +2 c_{2} \left( \frac{1}{4} g_{\mu \nu} W^{\lambda \rho} W_{\lambda \rho} -W_{\mu \lambda}W^\lambda_\nu \right)\nonumber \\
&& \left.  +c_{3} \left( \frac{1}{4} g_{\mu \nu} G^{\lambda \rho} G_{\lambda \rho} -G_{\mu \lambda}G^\lambda_\nu \right)\right],
\end{eqnarray}
where  $a=(1,2,3)$ denote the (U(1),SU(2),SU(3)) gauge fields in SM respectively, $F_{\lambda \rho, a}$ is the field strength tensor of gauge field, and the relations between the dimensionless coupling constants are
\begin{eqnarray}
c_{\gamma \gamma} &=& c_1 \cos^2\theta_W + c_2 \sin^2\theta_W,  \nonumber \\
c_{Z \gamma} &=& \left( c_2 - c_1 \right) \sin\left( 2\theta_W  \right), \nonumber \\
c_{Z Z} &=&c_1 \sin^2\theta_W + c_2 \cos^2\theta_W,  \nonumber
\end{eqnarray}
where $\theta_W$ is the Weinberg angle.
In this work, for simplicity we set $c_2 = c_1$ and then have $c_{\gamma \gamma}=c_{Z Z}=c_1$ and $c_{Z \gamma} = 0 $ (i.e., the channel
of $G \rightarrow Z \gamma $  is absent).

\subsection{Data}
\label{sec:constraints}

No
significant resonance is found in different channels by ATLAS and CMS, such as $W^+W^-$,$ZZ$ and so on.
Here we use the upper limits on cross sections in $W^+W^-$,$ZZ$, dijet and diphoton channels \cite{pp2ww,pp2zz,pp2jj,pp2aa} for spin-2 resonance with different masses to constrain the parameter space.
Here we use FeynRules \cite{Feynrules} and Madgraph \cite{madgraph} to calculate these cross sections.

The final states with a monojet and large missing
transverse momentum in $pp$ collisions are usually used to search dark matter signal. Here we use Madgraph \cite{madgraph} to simulate
this process and choose the same cuts as that adopted in Ref.\cite{constraintmonoj} and we use the signal region SR7 to get the constraints.

We also consider the constraints from the dark matter relic density and indirect detection results. The dark matter relic density from
the Planck Collaboration is $\Omega_{\rm DM}h^2 =0.1186 \pm 0.0020$.
Here we use a loose constraint, $0.09<\Omega_{\rm DM} h^2<0.13$ (calculated with micrOMEGAs \cite{micro}).
In this model, dark matter particles can annihilate into two photons with massive graviton as the mediator, which yields the $\gamma$-ray line that
is expected to be the smoking gun signal of the dark matter (see \cite{LiangYF2016} and the references therein). Hence we can constrain
the parameter space with the upper limits set by the $\gamma$-ray line search of Fermi-LAT \cite{Fermi2} and HESS. \cite{HESS}. Moreover, the
yielded continuous $\gamma$-ray emission is examined.

\begin{figure}[htbp]
\centering
\subfigure{
\includegraphics[width=0.43\textwidth]{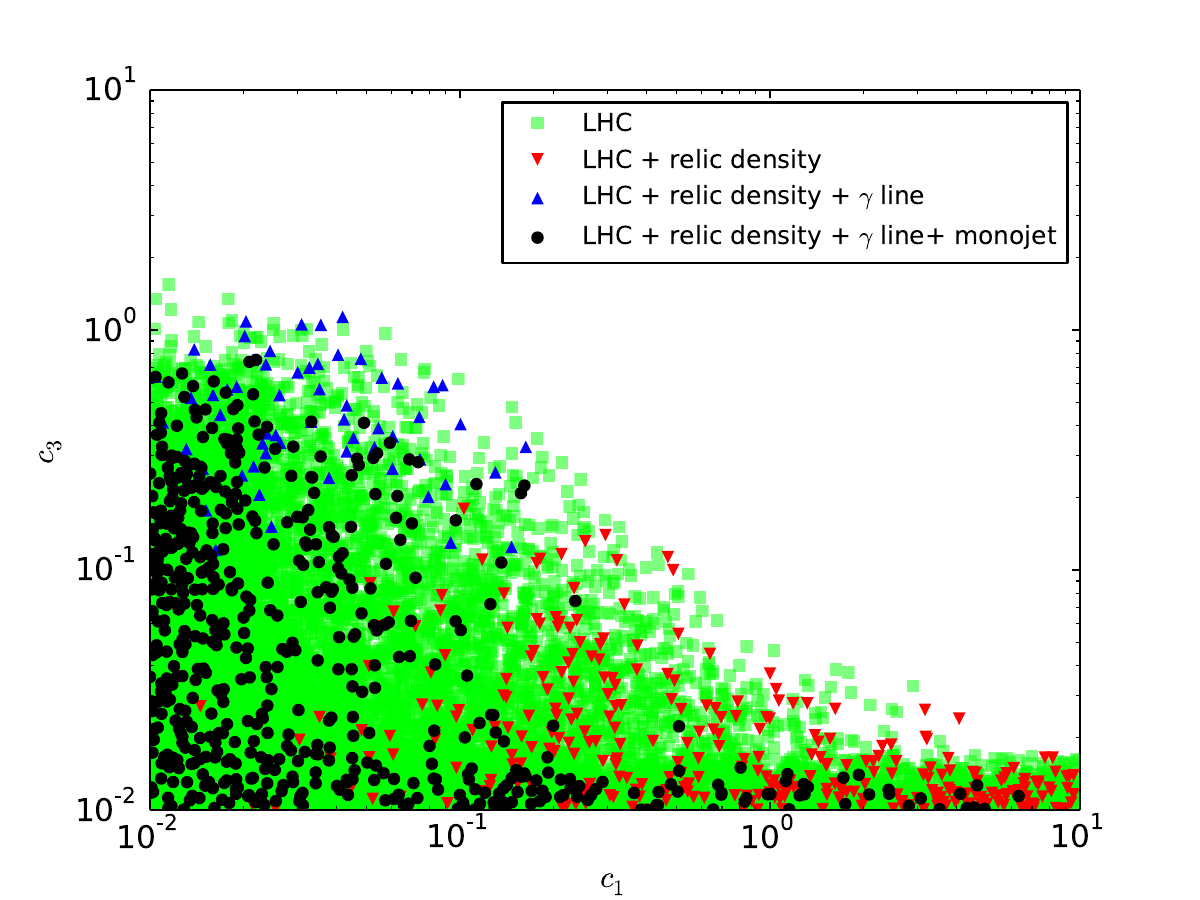}  }
\subfigure{
\includegraphics[width=0.43\textwidth]{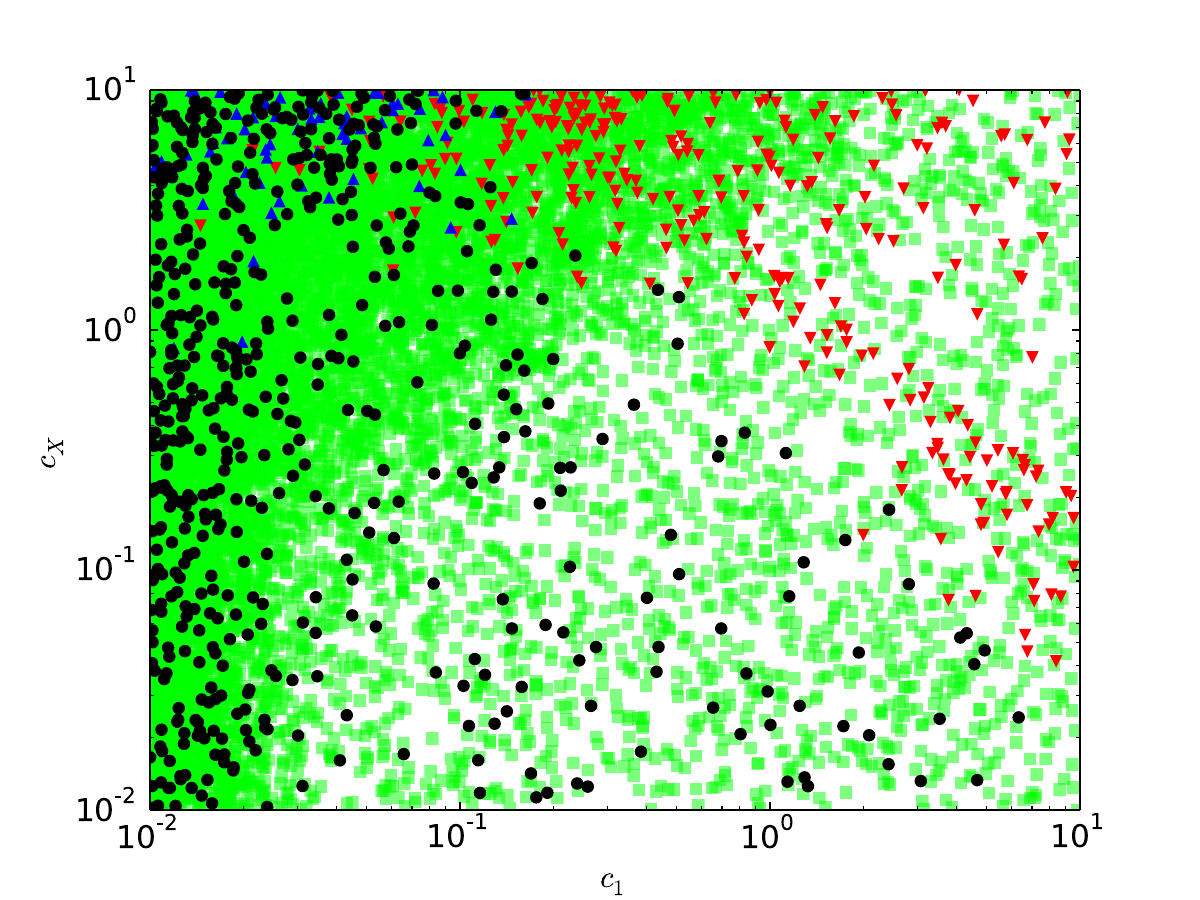}}
\subfigure{
\includegraphics[width=0.43\textwidth]{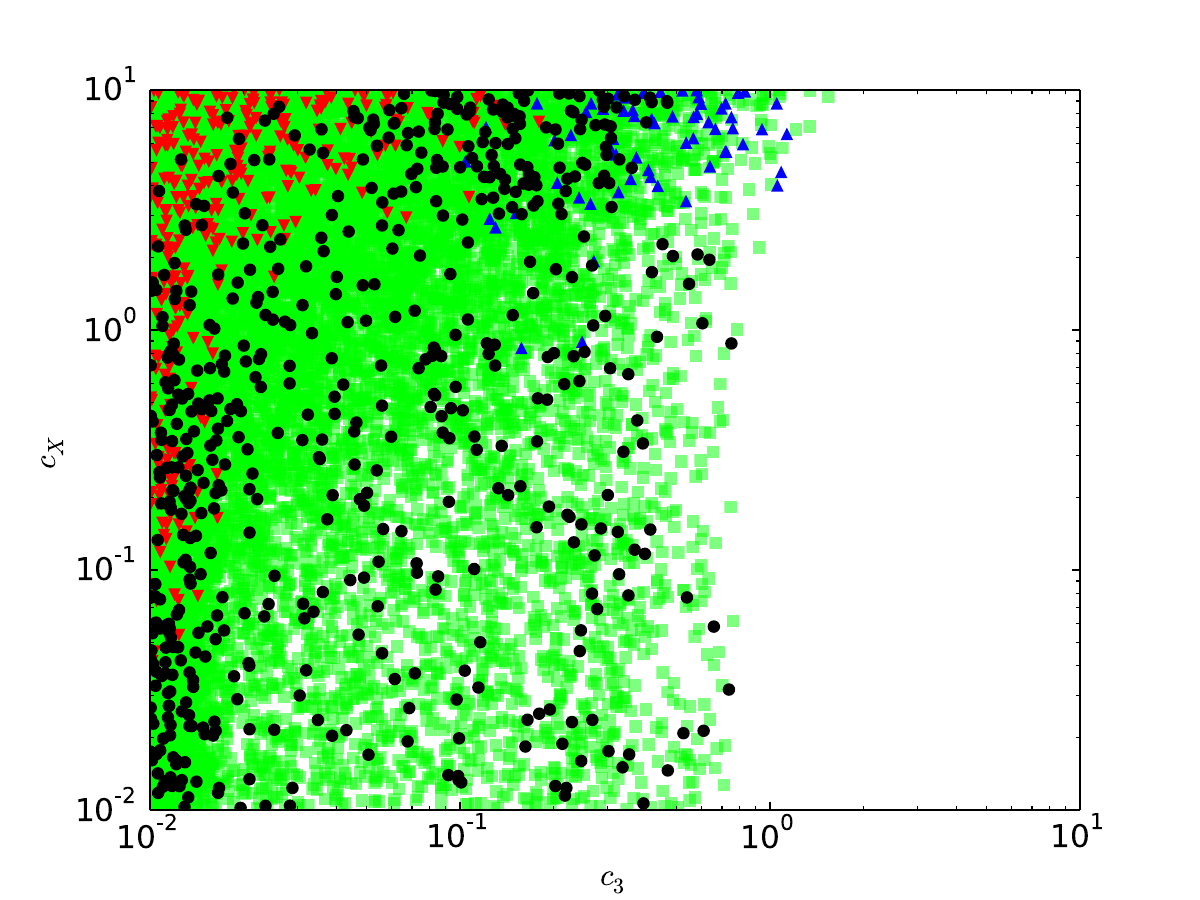} }
\subfigure{
\includegraphics[width=0.43\textwidth]{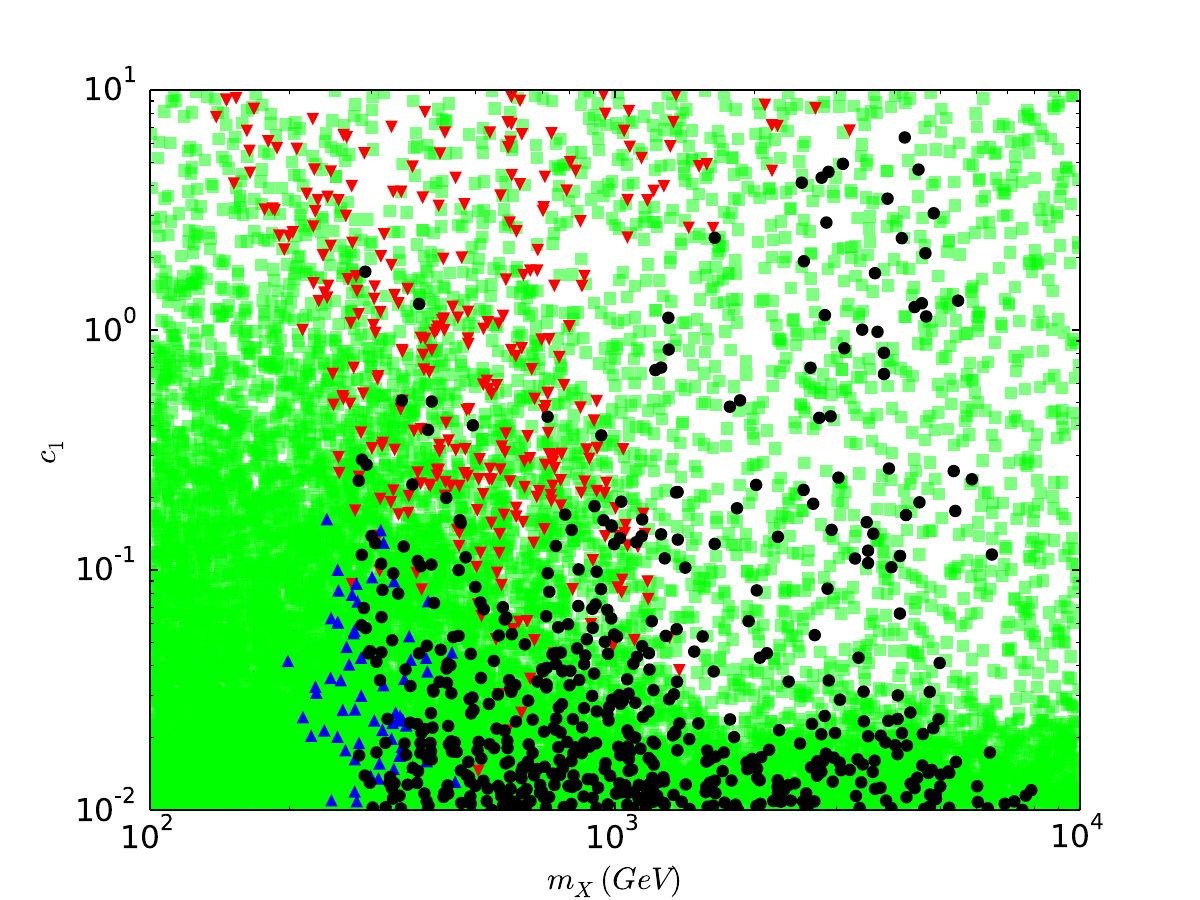} }
\subfigure{
\includegraphics[width=0.43\textwidth]{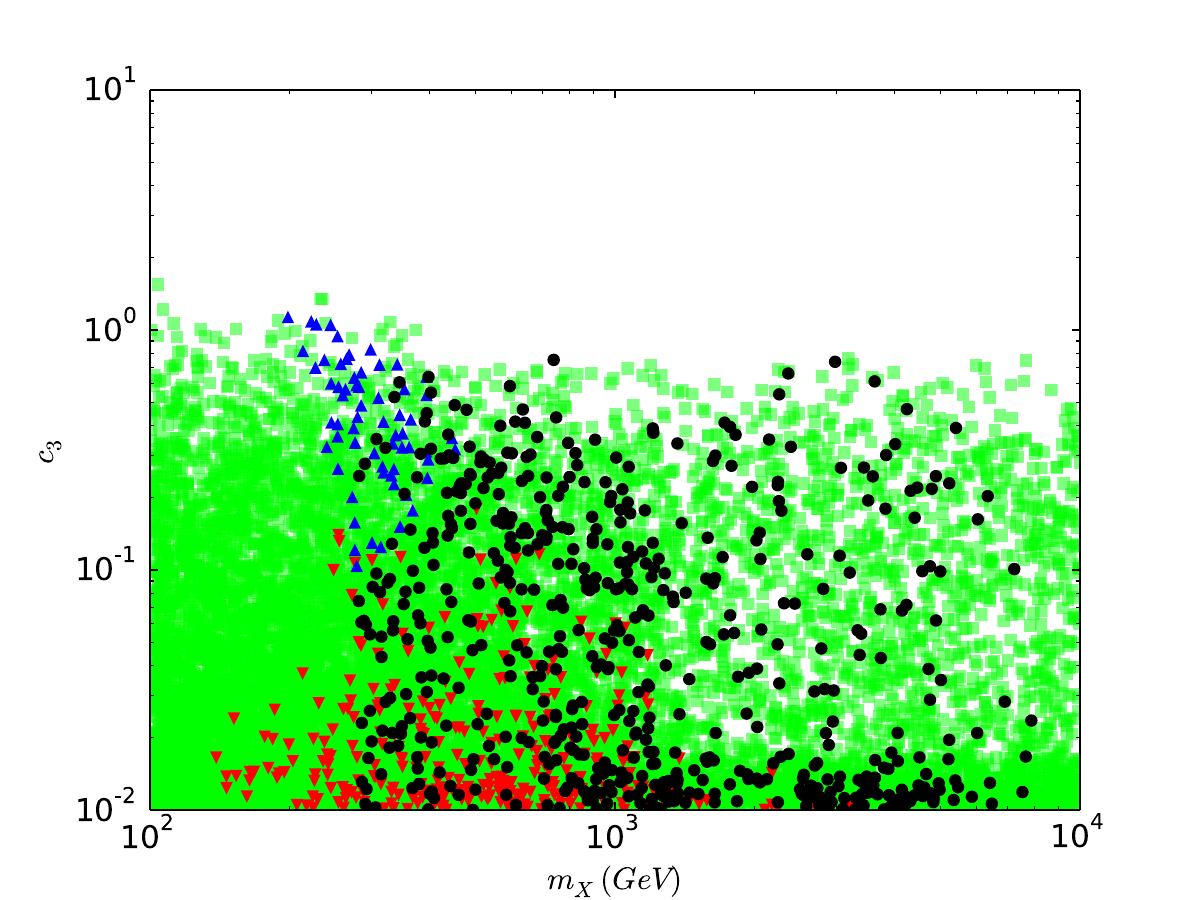} }
\subfigure{
\includegraphics[width=0.43\textwidth]{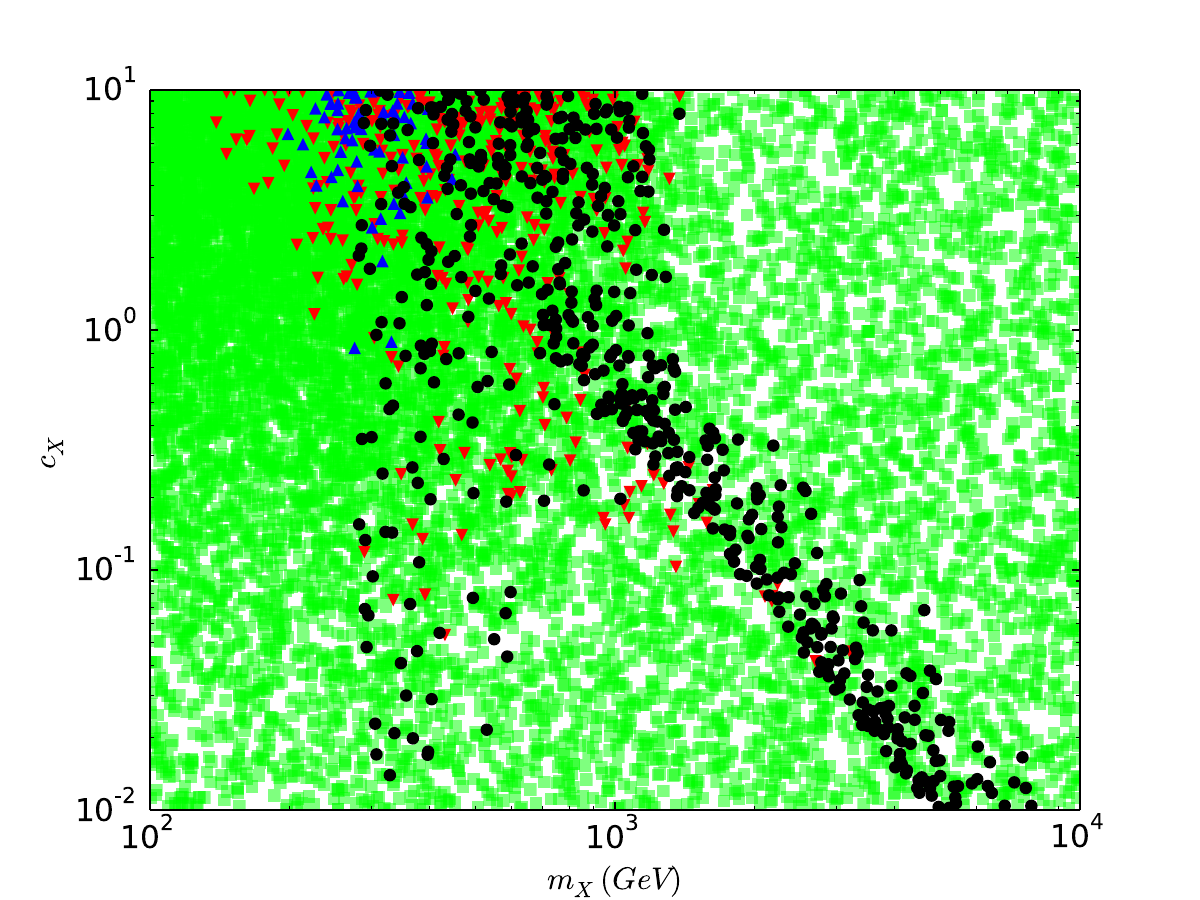} }
\caption{The constraints on the parameters of the vector dark matter model. The green points represent the allowed parameters permitted
by the LHC data without monojet constraints. The red (blue, black) points represent the parameter regions passing the further constraints
from the DM relic density data (DM relic density data $+$ $\gamma$-ray line data, DM relic density data $+$ $\gamma$-ray line data $+$ monojet
plus missing energy data), respectively.}\label{fig1}
\end{figure}

\begin{figure}[htbp]
\centering
\subfigure{
\includegraphics[width=0.43\textwidth]{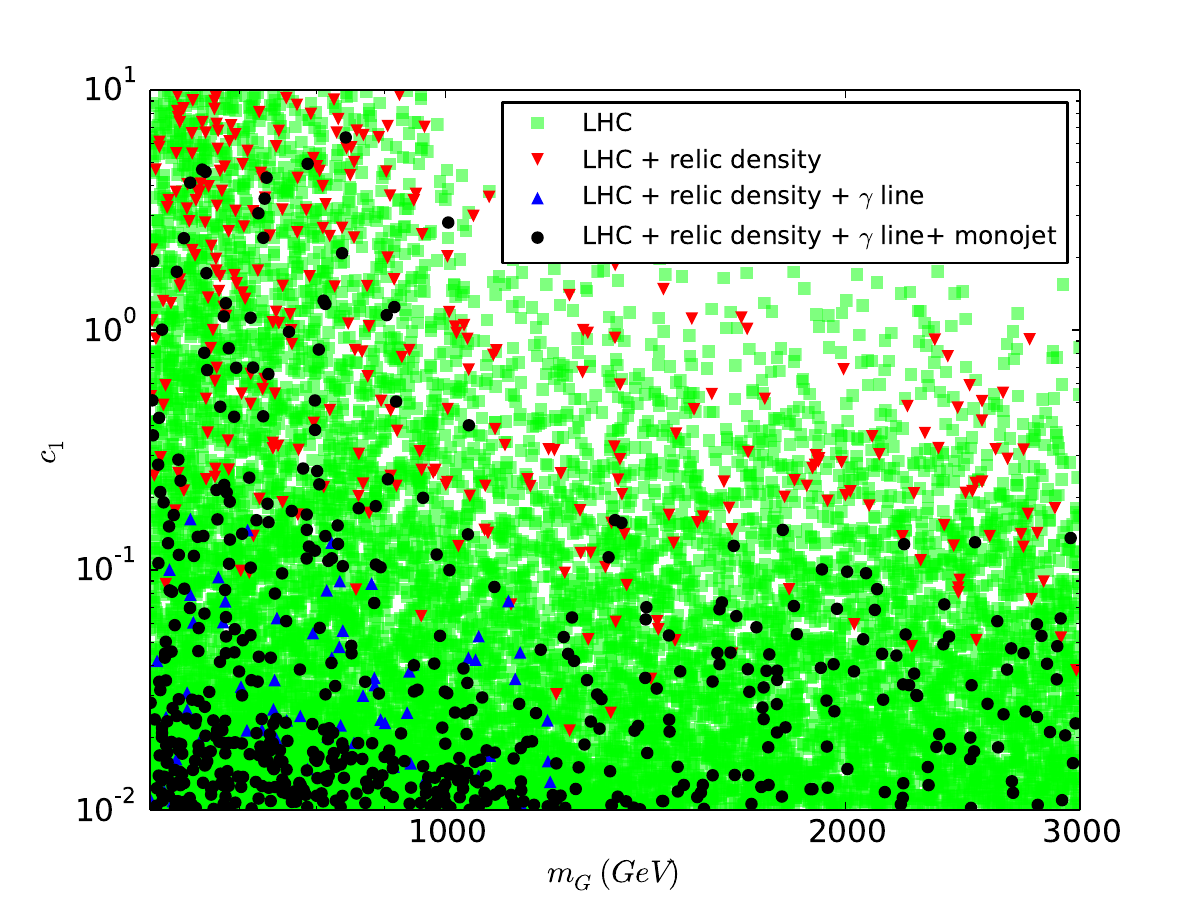} }
\subfigure{
\includegraphics[width=0.43\textwidth]{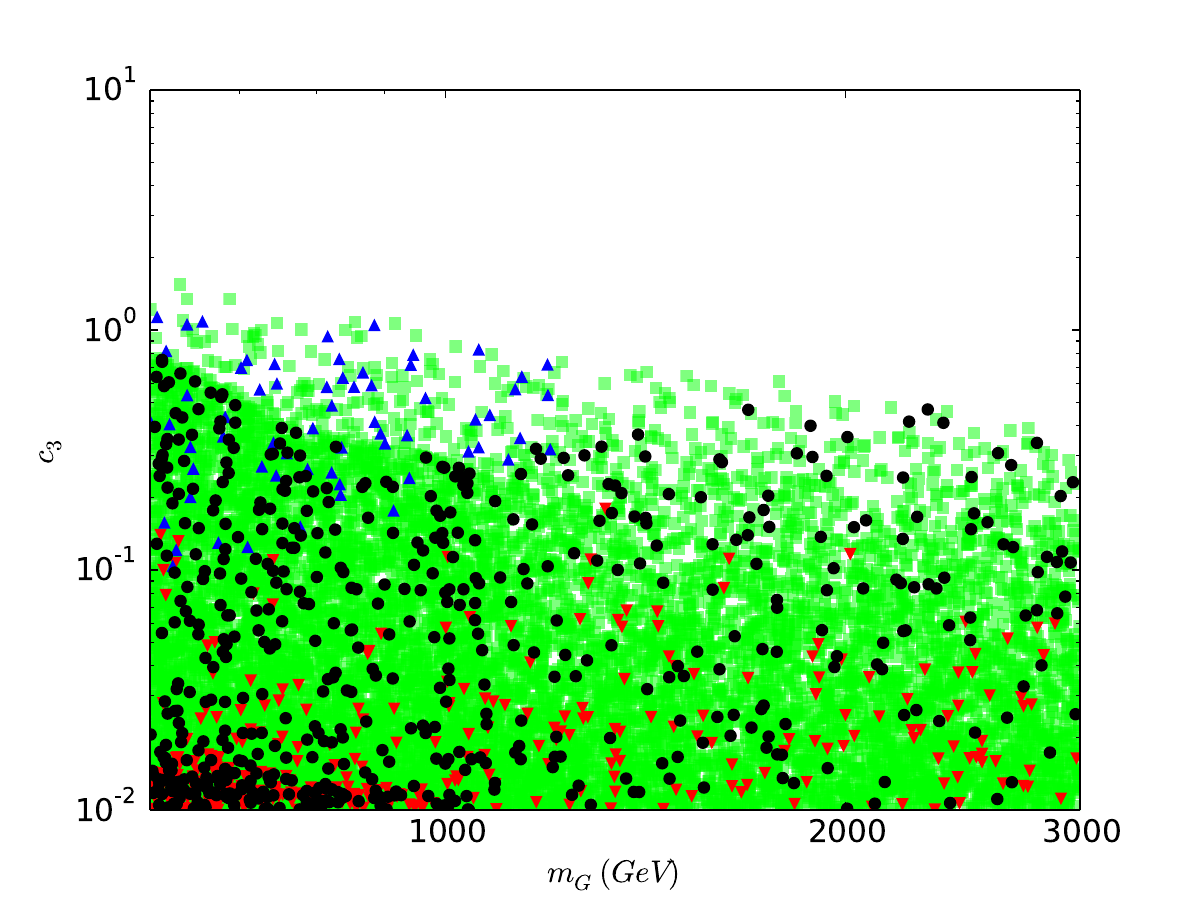} }
\subfigure{
\includegraphics[width=0.43\textwidth]{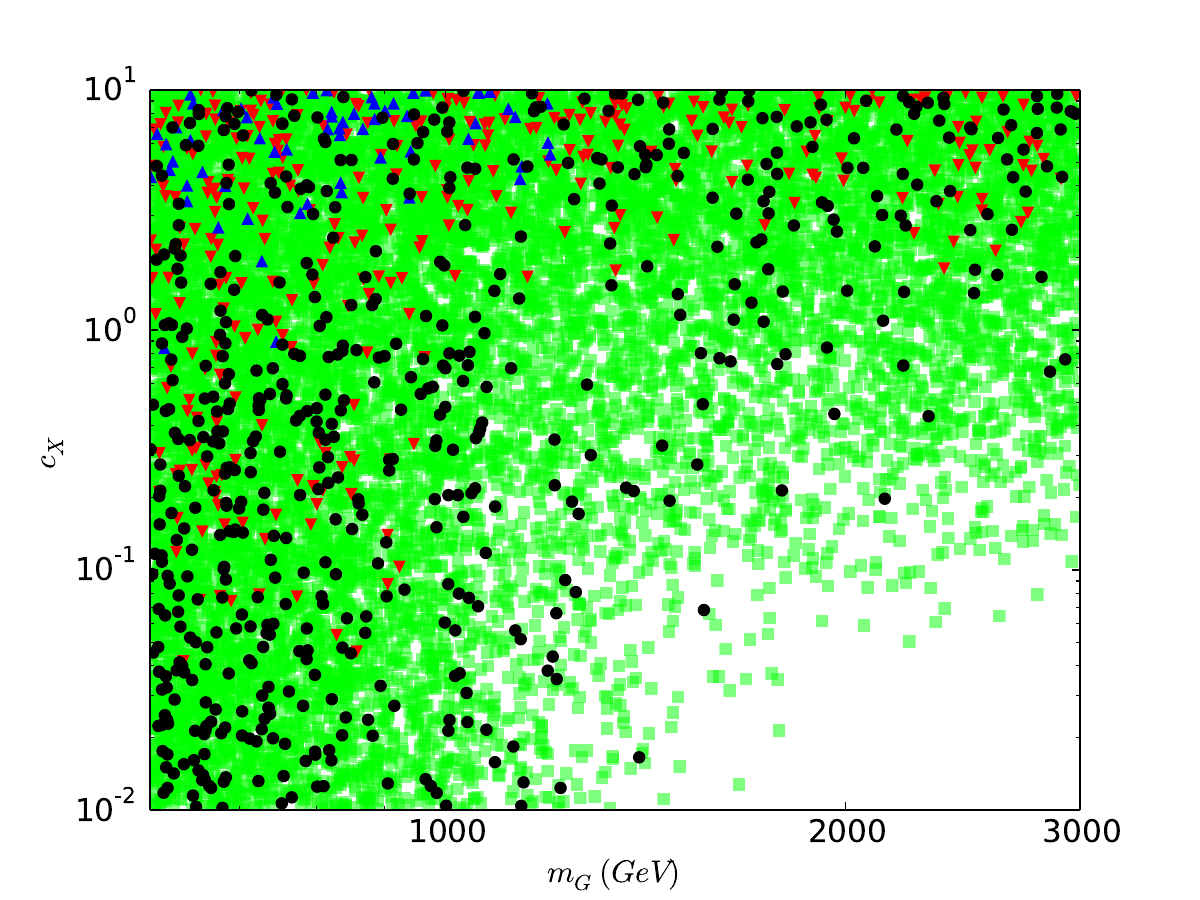} }
\subfigure{
\includegraphics[width=0.43\textwidth]{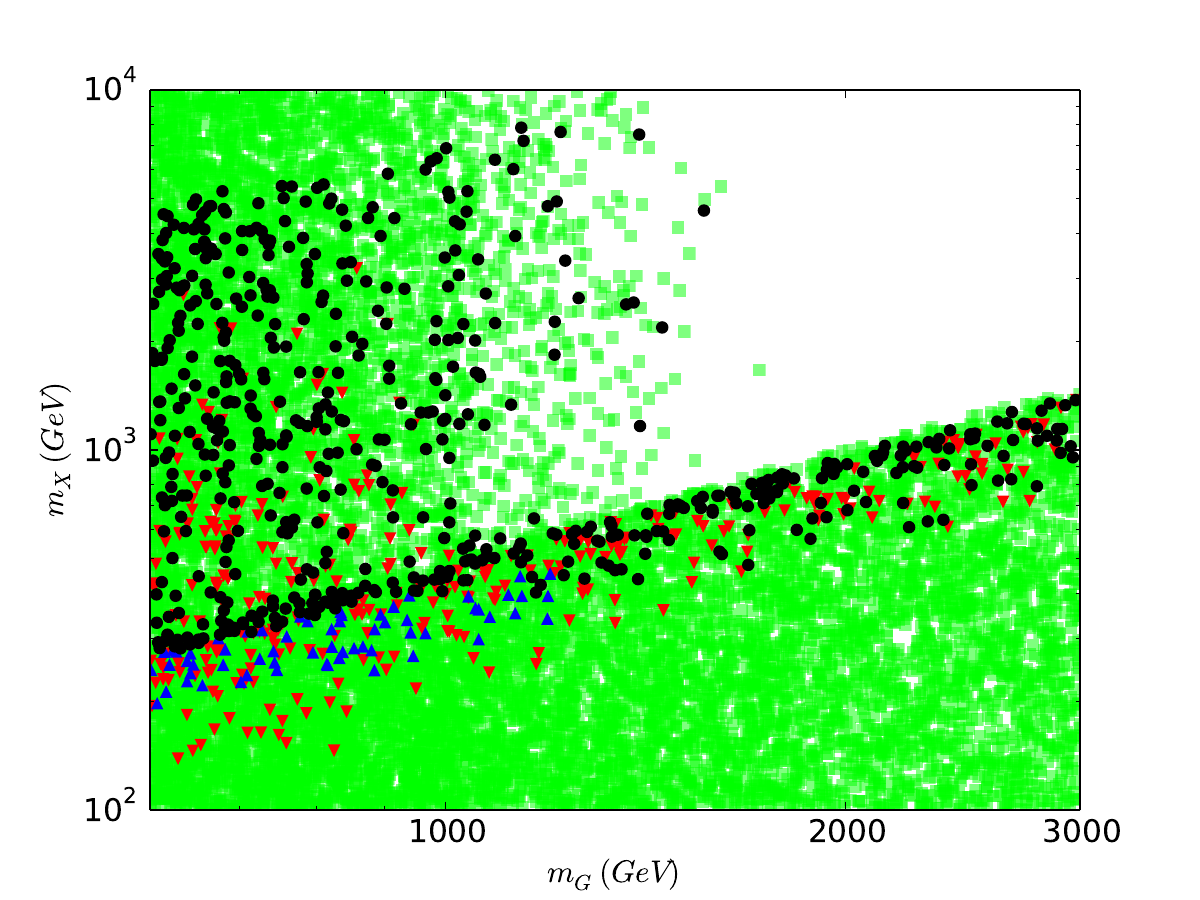} }
\caption{The constraints on the parameters of the vector dark matter model. The points are the same as described in Fig.\ref{fig1}.}\label{fig2}
\end{figure}

\subsection{Results}
 \label{sec:results}
In our survey the free parameters are allowed to vary in larger range, i.e.,
\begin{eqnarray}
&&-2<\log_{10}(c_1)<1, \nonumber \\
&&-2<\log_{10}(c_3)<1, \nonumber\\
&&-2<\log_{10}(c_X)<1, \nonumber \\
&&600<m_G/{\rm GeV}<3000, \nonumber \\
&&2<\log_{10}(m_X/{\rm GeV})<4, \nonumber
\end{eqnarray}
where $c_X$ is the dimensionless coupling constant between vector dark matter and massive graviton,  $m_G$ is the mass of the vector dark matter.
It should be pointed out that the results depend on the full coupling constant $c/\Lambda$ and in this work the coupling scale $\Lambda$ is set as 3 TeV. The range of $m_G$ is $\rm 600GeV \sim 3000GeV$ because in this range there are constraints on all the channels considered here. As there are no clearly signals on a spin-2 resonance, the coupling parameters can be very small or zero and then we set the lower bound of the coupling parameters as 0.01.
The constraints are applied in the following order: LHC upper limits, DM relic density, $\gamma$-ray line and LHC monojet search. Using this method, we finally get the sensitive range for each observations.

\begin{figure}[htbp]
\centering
\subfigure{
\includegraphics[width=0.46\textwidth]{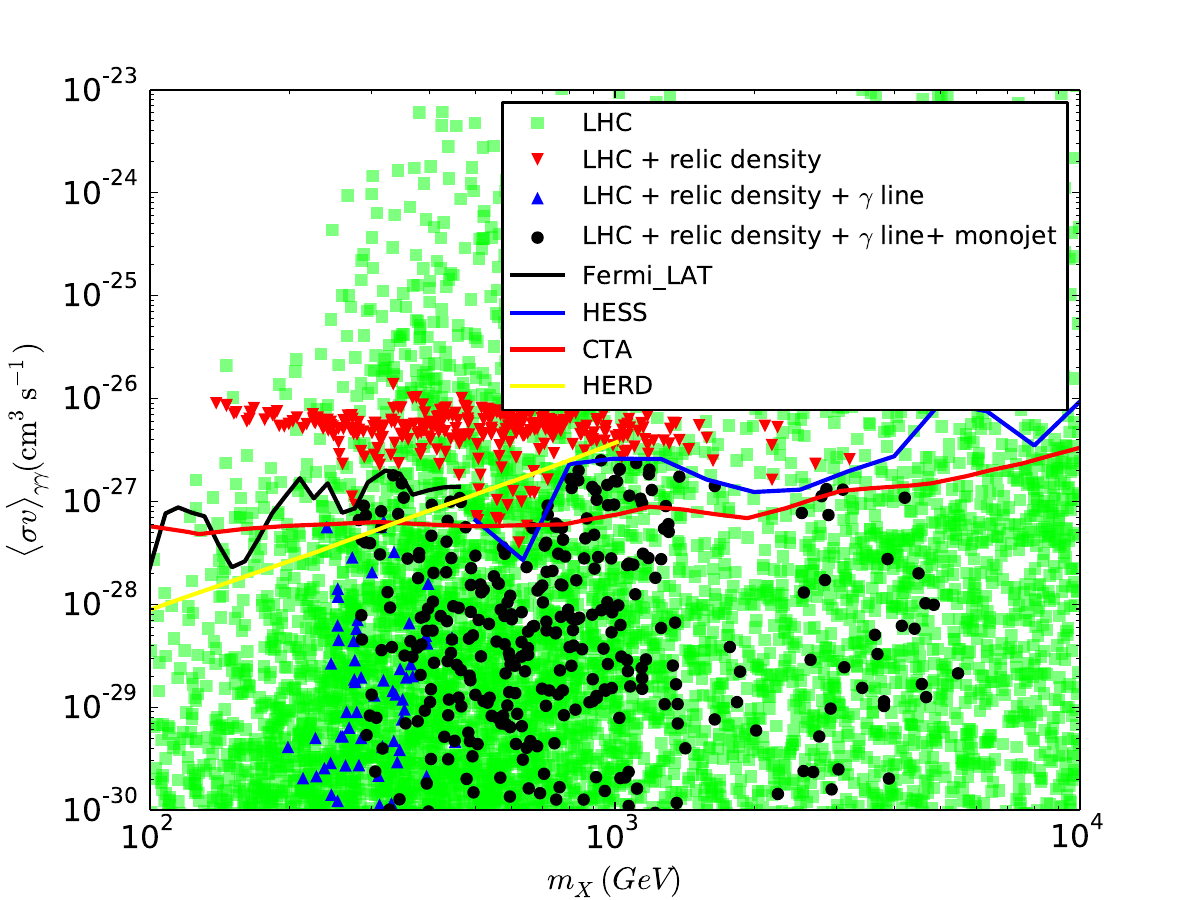}  }
\subfigure{
\includegraphics[width=0.46\textwidth]{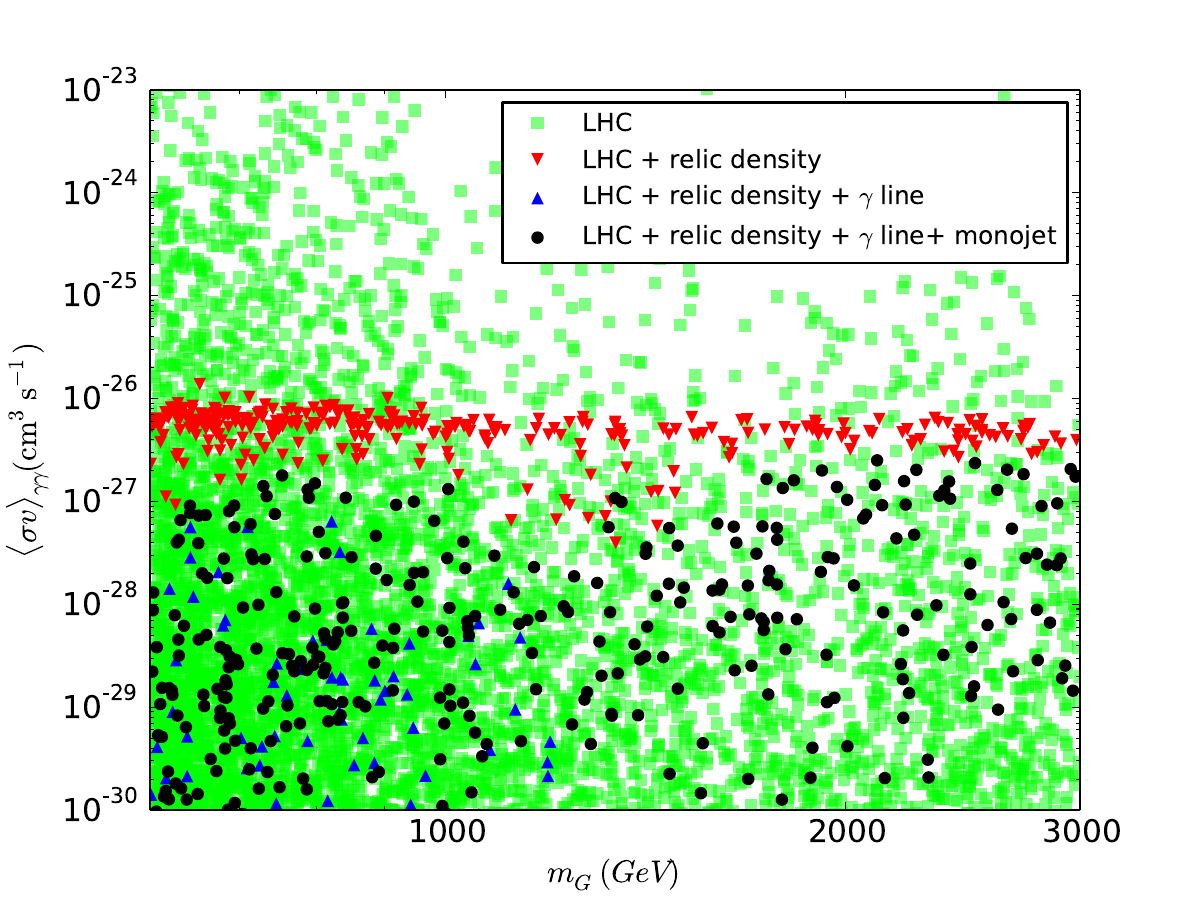}  }
\caption{Left: the scatter plots of the dark matter mass m$_X$ vs $<\sigma v>_{\gamma \gamma}$ the annihilation cross section of
channel $X X \rightarrow \gamma \gamma$; Right: the graviton mass m$_G$(right ) vs $<\sigma v>_{\gamma \gamma}$.
The black and blue lines are the upper limits
of $\gamma$-ray line set by the Fermi-LAT \cite{Fermi2} and HESS \cite{HESS} data, respectively. The red (yellow) line is the expected
line-search sensitivity of the CTA \cite{CTA} (HERD \cite{herd}).}\label{fig3}
\end{figure}

Fig.\ref{fig1} and Fig.\ref{fig2} show the survived parameters of each observation. It is straightforward to see that the relic density imposes a very tight constraint on the
parameters and just a small fraction of the points passing the LHC constraints (green points) is still allowed. Then the $\gamma$-ray line
and LHC missing
energy constraints are applied, and the black points are the parameters satisfied by all constraints. Points with large $c_1$ and $c_3$ have large annihilation cross section for
channel $X X \rightarrow \gamma \gamma$, and larger cross section $\langle \sigma v\rangle_{\gamma\gamma}$ have been rejected by the Fermi-LAT and HESS line-search data (please see the red points in the upper right panel of Fig.\ref{fig3}).
It is clear that the parameter space of $m_X$ and $m_G$ divided into two parts. For the part where $m_X$  is smaller than $m_G/2$, the massive graviton can decay to dark matter particles, and the dark matter decay ratios of the graviton are large for most of the black points.

In Fig.\ref{fig3} we show the allowed regions as a function of the cross section of the annihilation channel
$XX \rightarrow \gamma\gamma$ (i.e., $\langle \sigma v\rangle_{\gamma\gamma}$) and the mass of vector dark matter( massive graviton).
The constraints from Fermi-LAT \cite{Fermi2} and HESS \cite{HESS} are applied. The expected lines search sensitivity of the
Cherenkov Telescope Array (CTA)\cite{CTA},  the next generation ground-based very high energy $\gamma$-ray instrument that is under construction
\cite{CTAhttp}, and the High Energy cosmic-Radiation Detection(HERD)\cite{herd0,herd} proposed onboard China's space station are also plotted.
It is clear from the figure that the constraints from Fermi-LAT and HESS are
powerful constraints and reject a large region. And the CTA and HERD are expected to probe some regions
 (i.e., the black points above the red or yellow line).

\begin{figure}[htbp]
\centering
\subfigure{
\includegraphics[width=0.46\textwidth]{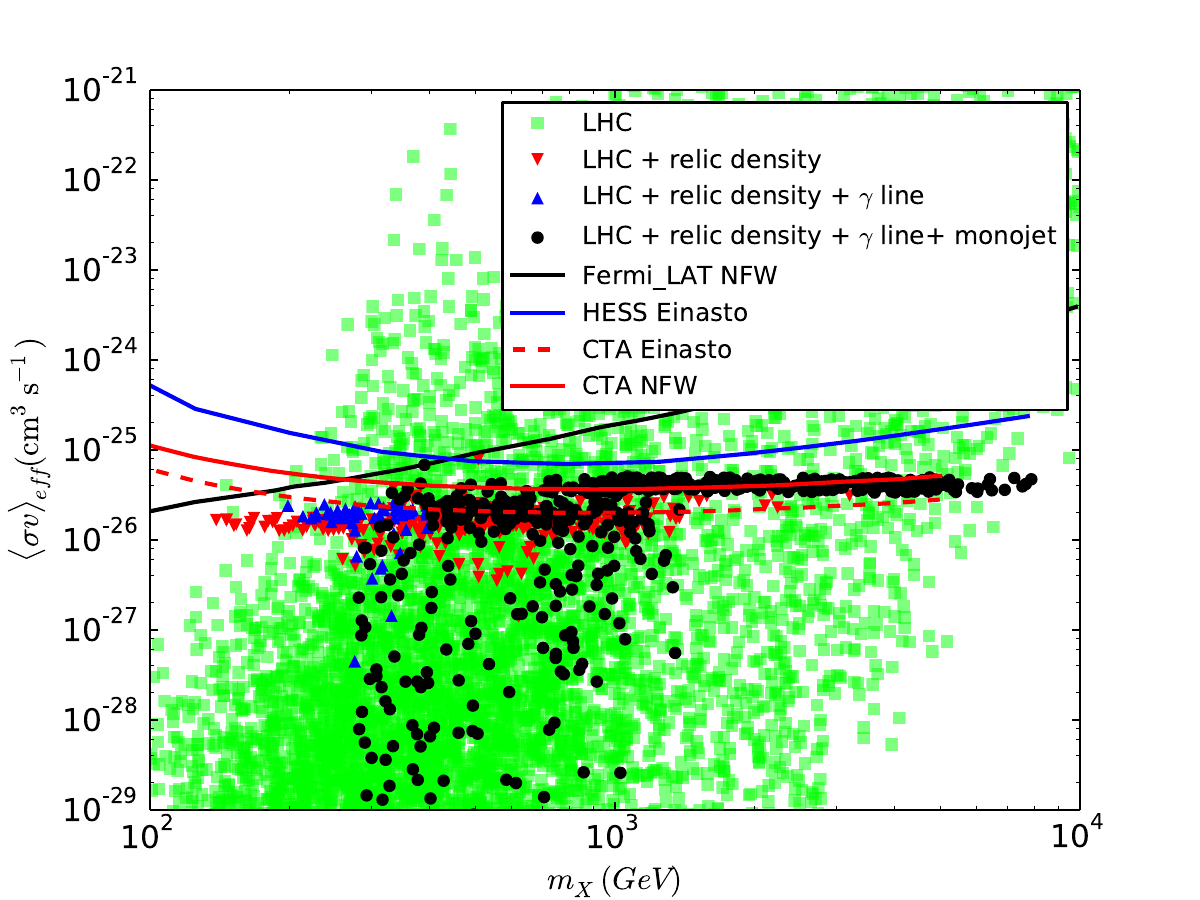}  }
\subfigure{
\includegraphics[width=0.46\textwidth]{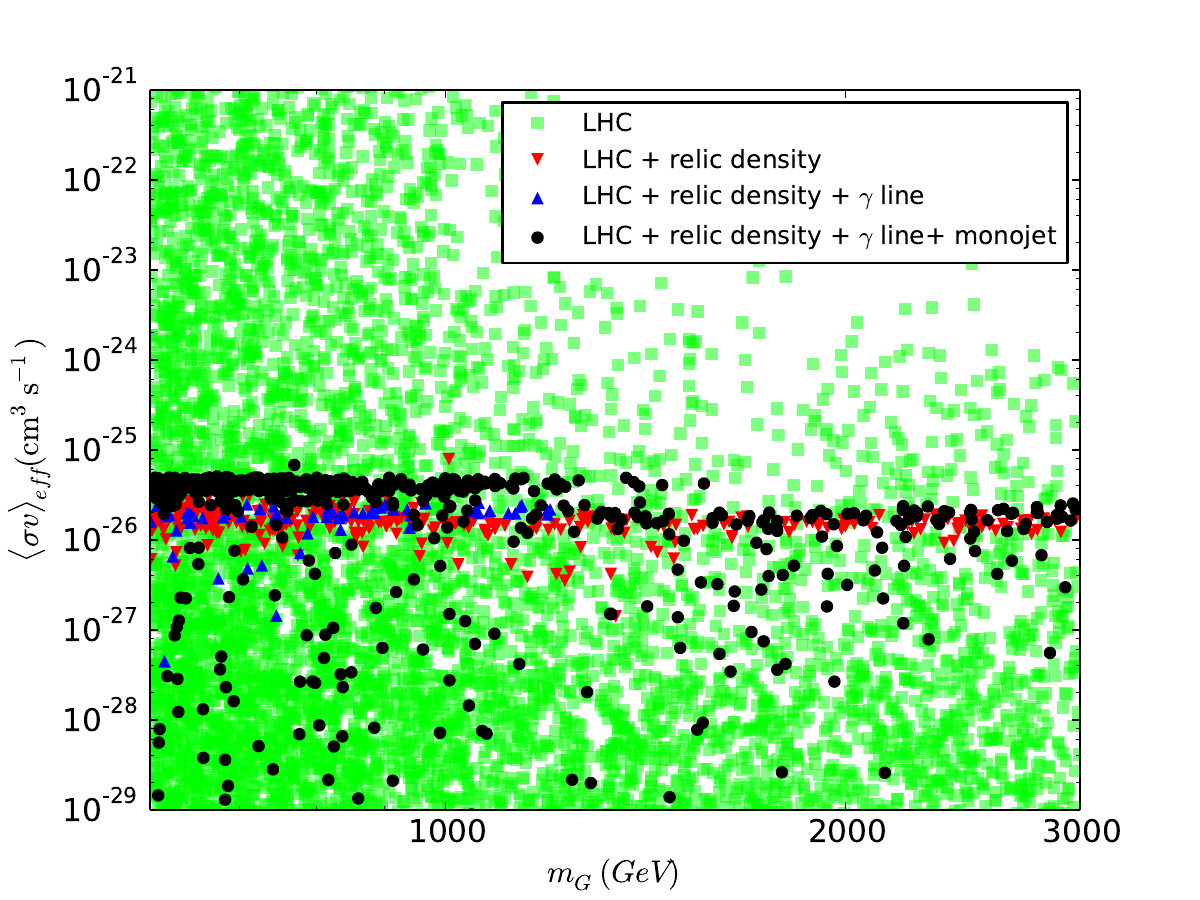}  }
\caption{Scatter plot of the dark matter mass m$_X$(left) and the graviton mass m$_G$(right) vs the effective cross section for continuous $\gamma$-ray emission. The black line  is the limit of the annihilation cross section for pure $b \bar{b}$ channel from  Fermi-LAT \cite{Fermi1} for Navarro-Frenk-White (NFW)  dark matter profile \cite{NFW}.
The blue line is the limits of the annihilation cross sections for pure $b \bar{b}$ channel from HESS \cite{HESS-bb} for Einasto dark matter profile
\cite{Einasto}.
The red line (dash line) is the expected sensitivity of the CTA \cite{CTA1} for NFW(Einasto) profile.}\label{fig4}
\end{figure}

Besides the $\gamma$-ray line, the continuous $\gamma$-ray would be produced if dark matter particles annihilates to $gg$, $ZZ$, $W^+ W^-$ and $GG$. Following \cite{bi2015}, we calculate the effective total cross section in order to compare with observations, i.e.,
\begin{eqnarray}
\left \langle \sigma v  \right \rangle_{eff} = \left \langle \sigma v  \right \rangle_{ZZ}+\left \langle \sigma v  \right \rangle_{W_+ W_-}+
\left \langle \sigma v  \right \rangle_{gg}+2
\left \langle \sigma v  \right \rangle_{GG}.
\end{eqnarray}
Considering the similarity of initial $\gamma$-ray spectra produced by the $gg$, $b\bar b$ and light quark final states, the corresponding upper limits of these final state are also similar with each other. Thus the effective total cross section can be compared with upper limitations for pure channel roughly.
Fig.\ref{fig4} shows the effective total cross section comparing with the upper limits of dark matter annihilation cross sections for
pure  $b \bar{b} $ channel from Fermi-LAT \cite{Fermi1} and HESS \cite{HESS-bb}. The expected line search sensitivity of CTA is also plotted. For $m_X>{\rm 1500~GeV}$ region where the corresponding $m_G$ is smaller than {\rm 1500~GeV}(Fig.\ref{fig2}), the dark matter particles can annihilate into
two gravitons which makes the effective cross sections slightly larger. For dark matter mass larger than $1500~{\rm GeV}$, the dark matter
particles mainly annihilate to gravitons, and the effective cross section approximately equals to $3\times10^{-26}~{\rm cm^3 ~ s^{-1}}$.
In two regions the effective cross section can vary in a large scale.
The first region is near the resonance of DM s-channel annihilation to SM particles where $m_X = m_G/2$.
And the second region is near the threshold of t-channel annihilation to massive graviton pair where the t-channel annihilation, dominating in the calculation of DM relic density, contributes little to the effective cross section.
Therefore for region with $200~{\rm GeV}<m_X<1500~{\rm GeV}$ which have points with $m_X \approx m_G/2$ (the resonance region) or $m_X \approx m_G$ (the threshold of t-channel annihilation to massive graviton pair), the effective cross section varies in a large scale.
Due to their limited sensitivities, current $\gamma$-ray experiments such as Fermi and
HESS can not effectively probe the allowed parameter space, yet. But the situation will change soon since the next generation experiment CTA would effectively constrain the region $m_X>1500~{\rm GeV}$. The effective cross sections change little for different values of $m_G$ (right part of Fig.\ref{fig4}).

\subsection{The 750 GeV diphoton excess}
 \label{sec:750}

 \begin{figure}[htbp]
\centering
\subfigure{
\includegraphics[width=0.46\textwidth]{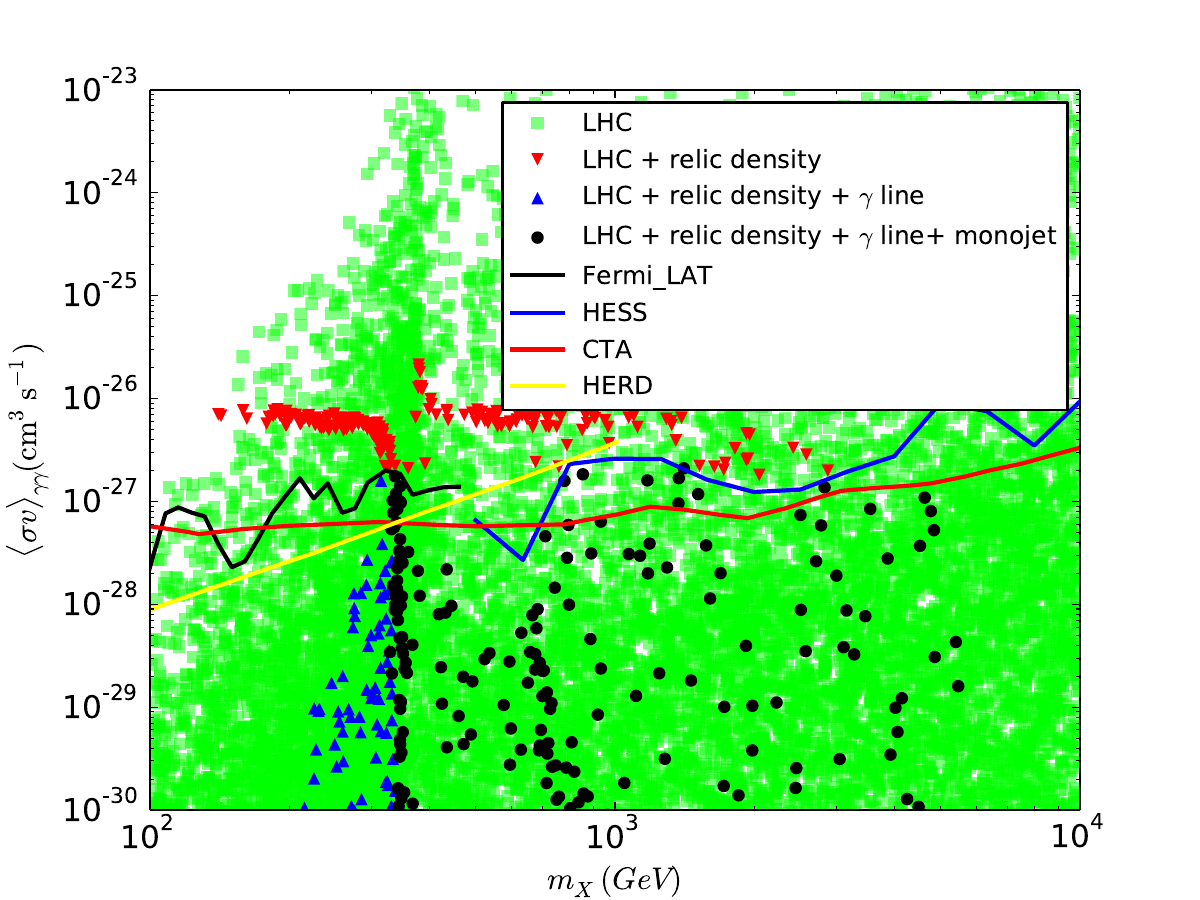}  }
\subfigure{
\includegraphics[width=0.46\textwidth]{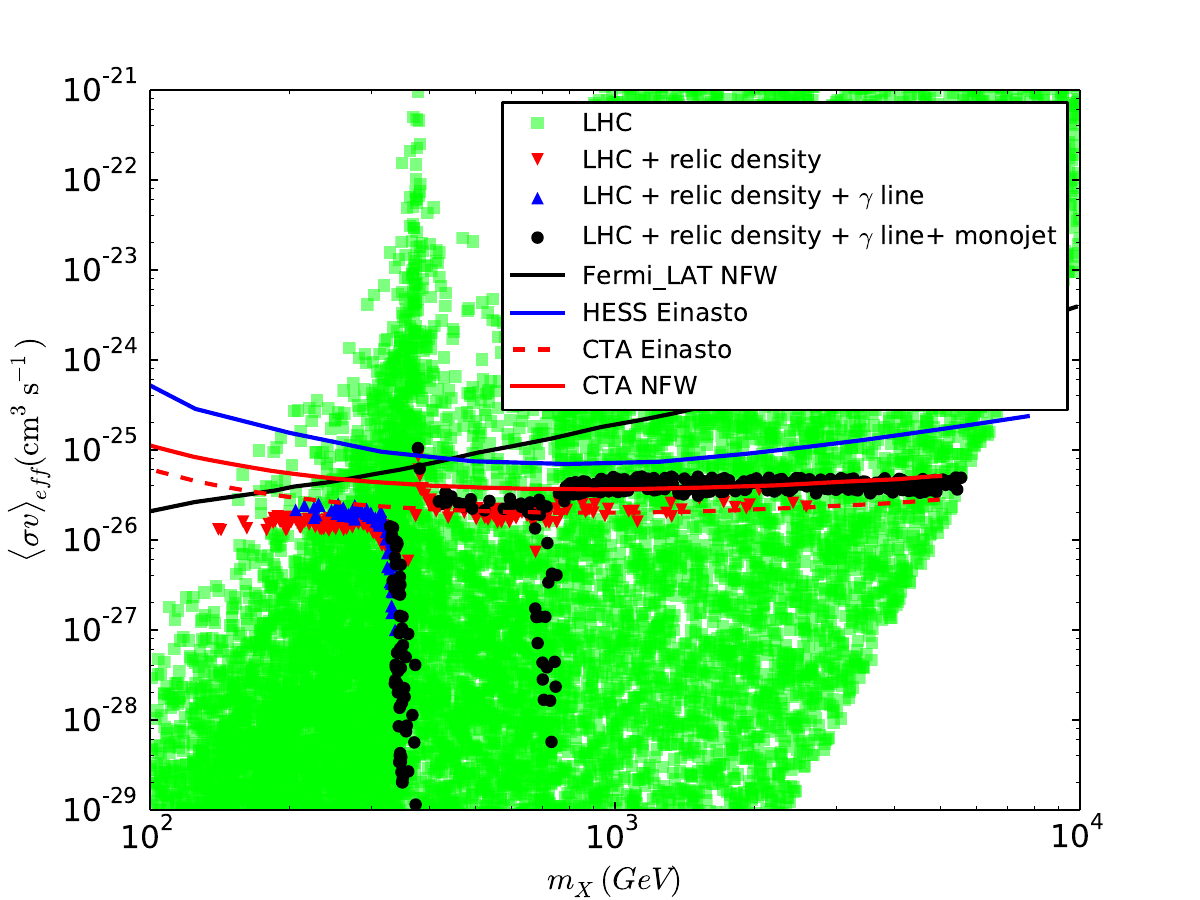}  }
\caption{Scatter plot of the dark matter mass m$_X$ vs $<\sigma v>_{\gamma \gamma}$(left) and the effective cross section for continuous $\gamma$-ray emission (right) with the graviton mass fixed as 750 GeV. }\label{fig5}
\end{figure}
In this work, we set the massive graviton mass to be free. If the massive graviton has been detected, the mass is fixed (for example, in the graviton interpretation of the diphoton excess, the mass should be 750 GeV).  The 750 GeV diphoton excess was reported by CMS and ATLAS in the first results of LHC run-2 in 2015 \cite{ATLAS01,CMS01} and was
considered as a indication a new particle. While with more data collected by LHC, no 750 GeV diphoton excess was found by CMS for both spin-0 and spin-2 resonances. ATLAS reported the null observation of spin-0 resonance (for spin-2 resonance the data has not been published, yet). In view of these facts here we briefly discuss the possible $\gamma$-ray
signal for spin-2 massive graviton resonance with for example $m_G=750 {\rm GeV}$.

In Fig.\ref{fig5} we show the allowed regions of the cross section of the annihilation channel $XX \rightarrow \gamma\gamma$ (the left panel) and the effective cross section for continuous $\gamma$-ray emission (the right panel). The constraints from Fermi-LAT and HESS are applied.
The expected sensitivity of the CTA is also presented.  The annihilation cross sections $\langle\sigma v\rangle_{\gamma \gamma}$ (black points) vary from a pretty small number to a reasonable large value of $10^{-27}~{\rm cm^3~s^{-1}}$. CTA is expected to probe some regions (i.e., the black points
that are near the resonance). However, for $m_X>400~{\rm GeV}$ the annihilation cross sections are still too small to
be reliably tested. It is clear from the right of Fig.\ref{fig5} that effective total cross sections vary in a large scale at the region where $m_X \approx m_G/2$ (the resonance region) or $m_X \approx m_G$ (the threshold of t-channel annihilation to massive graviton pair).
And this region can not be probed well by the next generation experiment CTA.

\section{Summary}
\label{sec:summary}
In this work we studied the $\gamma$-ray signal in the massive graviton mediated dark matter model where the dark matter interact with SM gauge bosons through the exchange of massive graviton. We randomly generated
the model parameters and constrain them using LHC data, thermal relic density, $\gamma$-ray line data (upper limits) and continuous $\gamma$-ray data (upper limits) from Fermi-LAT and HESS.  For our purpose we
focused on vector dark matter for which the annihilation cross section of $\chi \chi\rightarrow \gamma \gamma$ is not suppressed, which is important for dark matter indirect detection since no other known physical processes can yield lines in GeV-TeV energy range and hence the successful detection will be taken as the discovery of dark matter
particles. Indeed, one of the leading scientific objectives of the space telescopes with high energy resolution (such as DAMPE, CALET and HERD), and the ground-based CTA is to
catch or impose stringent constraint on such a signal. The main results of this work are the allowed parameter regions(shown in Fig.\ref{fig1} and Fig.\ref{fig2}) that
pass all current constraints.  We found that the $\gamma$-ray line data played an important role in narrowing down the model parameter space.  The viable velocity-averaged cross
section of annihilation into $\gamma$-ray line can be (partly) probed by the upcoming ground-based CTA and proposed space telescope HERD.
For the continuous $\gamma$-ray emission, the total effective annihilation cross sections vary in a large scale for the regions with $m_X \approx m_G/2$ (the resonance region) or $m_X \approx m_G$ (the threshold of t-channel annihilation to massive graviton pair).
For other regions, the total effective annihilation cross sections are  $\sim 10^{-26}
{\rm cm^{3}~s^{-1}}$, such cross sections are consistent with current observational data and will be reliably probed by the upcoming CTA and HERD. The future results of those experiments will help to study this massive graviton mediated dark matter model.

\section*{Acknowledgement}
We thank Dr. Lei Wu for helpful discussion and suggestions. This work was supported by the 973 National Major State Basic Research and Development of China (Grant No. 2013CB834400, and No. 2013CB837000),
by the National Natural Science Foundation of China (Grant Nos. 11535004, 11375086, 11120101005, 11175085, 11235001, 11303096),  by the Science and Technology Development Fund of Macau under Grant No. 068/2011/A, the Youth Innovation Promotion Association
CAS (Grant No. 2016288) and the Natural Science Foundation of Jiangsu Province (Grant No. BK20151608).
\\

\end{document}